\documentclass{elsart}
\usepackage{graphicx,amssymb}
\journal{JMMM}

\newcommand{\be}{\begin{equation}}
\newcommand{\ee}{\end{equation}}
\newcommand{\comment}[1]{}

\begin{document}
\begin{frontmatter}

\title{Single-domain versus two-domain configuration in thin ferromagnetic prisms}
\author{Maria Gloria Pini},
\ead{mariagloria.pini@isc.cnr.it}
\author{Paolo Politi}
\ead{paolo.politi@isc.cnr.it}

\address{Istituto dei Sistemi Complessi, Consiglio Nazionale
delle Ricerche, Via Madonna del Piano 10, 50019 Sesto Fiorentino, Italy}

\begin{abstract}
Thin ferromagnetic elements in the form of rectangular prisms are
theoretically investigated in order to study the transition from
single-domain to two-domain state, with changing the in-plane
aspect ratio $p$.  We address two main questions: first, how general is the
transition; second, how the critical value $p_c$ depends on the
physical parameters. We use two complementary methods:
discrete-lattice calculations and a micromagnetic continuum
approach. Ultrathin films do not appear to split in
two domains. Instead, thicker films may undergo the above
transition. We have used the continuum approach to analyze recent
Magnetic Force Microscopy observations in $30$nm-thick patterned
Permalloy elements, finding a good agreement for $p_c$.
\end{abstract}


\end{frontmatter}


\section{Introduction}
\label{intro}

In recent years, arrays of patterned ferromagnetic dots have
received considerable interest owing to their possible
applications in high-density magnetic data storage \cite{pescia}
and spin-electronic \cite{prinz} devices, as well as for realizing
logic functionality \cite{imre}. High-resolution electron beam
lithographic techniques \cite{broers} are commonly used to
fabricate the samples, in such a way that all the particles in the
array are virtually identical to each other. As a consequence,
the measured properties of the array can be interpreted as the
individual properties of a single dot, provided that the dots are
sufficiently far spaced to neglect the interdot magnetostatic
interaction. In this way, using high-sensitivity magneto-optical magnetometry
techniques, the variation of the properties with
the shape in magnetic nanoelements could be experimentally
investigated for dots as thin as 3 nm \cite{cowburn}.
Single dot properties can also be investigated by magnetic force microscopy (MFM)
techniques~\cite{mic_tech}, which allow to focus on a single element.

In this paper, we will consider in-plane magnetized dots
\cite{pescia}. Within the plane, the preferred direction of the
magnetization is determined by a balance between
magnetic dipole-dipole interaction and
magnetocrystalline anisotropy~\cite{EPJB}. In Permalloy
elongated nanodots, the direction of the uniform magnetization
is determined by the shape-induced magnetic anisotropy,
since the crystalline anisotropy of the material is negligible.

In a recent paper~\cite{PPS} we studied rectangular monolayers of planar
spins located
on the sites of a two-dimensional triangular lattice and interacting via magnetic dipole-dipole
interaction only. For this lattice geometry, the infinite monolayer (ML)
is believed~\cite{ground_state} to have a ferromagnetic (FM) ground state,
which is degenerate with the orientation.
The infinite square lattice is instead believed to have a
microvortex ground state, which is degenerate with a local orientation
angle.
For specific values of this angle, we get a state of FM
lines antiferromagnetically coupled. Therefore, the difference between the
two types of lattice is often understood in terms of the coupling
between FM lines of spins, which is ferromagnetic in the triangular lattice
and antiferromagnetic in the square lattice.

In our paper~\cite{PPS},  rectangular monolayers were studied with
varying their size and aspect ratio $p$, defined as the ratio
between the smaller and the larger side of the
rectangle.\footnote{Such a definition was adopted here for better
convenience as regards the comparison with the experimental data
in Ref.~\protect\cite{imre}. Note that it differs both from the
one adopted by Aharoni in Ref.~\protect\cite{aharoni_jap} and from
the one we used in Ref.~\protect\cite{PPS} ($r=1/p$ rather than
$p$).} Two main results came out. First, passing from
rectangular, elongated finite systems to square-shaped ones, we
obtained clear evidence for a transition from a FM to a two-domain
and eventually to a macrovortex state configuration. Second, in
the thermodynamic limit, we found that the macrovortex seems to be
the lowest energy state whatever is the aspect ratio, but for
elongated samples the size must attain unphysically large values
to display such a state.

In this paper, we want to study the effect of a direct exchange
interaction between spins, taking into account
the effect of a variable thickness as well.
These generalizations make our result relevant for FM prisms
which are currently available.
A recent paper~\cite{allenspach} by Jubert and Allenspach
moved along similar lines, with the authors studying
the transition from single-domain to macrovortex configuration
for a circular ferromagnetic disk.
Here, we are rather interested in the problem of the transition from
single-domain to two-domain configuration. For this reason,
we study a thin rectangular ferromagnetic prism.

We want to address two main questions:

i)~How general is the transition? We have already shown \cite{PPS}
that this transition occurs for a purely dipolar system in the ML
limit. Introducing the exchange coupling, $A$, makes the domain wall
more energetically expensive and change the width of the domain
wall: the latter is atomically sharp for a purely dipolar system,
while it spreads over many lattice constants when $A\ne 0$. Here,
we want to ascertain whether these changements are able to cancel the
transition and whether lattice structure effects, which are so
important for a purely dipolar system, are maintained.

ii)~If a transition occurs, how does the critical value,
$p_c$, of the in-plane aspect ratio, $p$,
of the rectangular element depend on the physical parameters
of the system?

We will use two complementary methods for the evaluation
of the magnetostatic energy difference, $\Delta E_M$, between a
two-domain state and a single-domain state:
discrete-lattice calculations, which are valid for ultrathin films,
and a micromagnetic continuum approach, applicable to thicker films.

In Section II discrete-lattice calculations are performed for a
rectangular monolayer, both for spins located on a triangular and
on a square lattice. Our main result is that the transition
from a single-domain state to a two-domain state is suppressed,
in the ultrathin limit, because of the exchange-induced
broadening of the domain wall, which modifies the energetic balance
of the dipolar energy. This result does not depend on the type of lattice,
because spin configurations varying on scales much larger
than the lattice constant depend weakly on lattice geometry.

In the case of thicker samples, a micromagnetic approach
is appropriate. In Section III we calculate the
magnetostatic energy difference, $\Delta E_M$, between the
two-domain and the single-domain state, following a method devised
a few years ago by Aharoni~\cite{aharoni_jap}. In the limit of
domain-wall width much smaller than the lateral dimensions of the
ferromagnetic dot, analytical expressions can be obtained both for
the magnetostatic energy gain, $\Delta E_M$, due to {\it surface}
charges and for the energy cost (per unit wall area), $\gamma_N$,
of a one dimensional N\'eel wall
\cite{aharoni_neel_wall,aharoni_book}. For a thin rectangular
prism, $\gamma_N$ includes contributions from the exchange
interaction, the uniaxial magnetocrystalline anisotropy, and the
magnetostatic energy due to {\it volume} charges in the N\'eel
wall. One can thus estimate in a simple way the critical value,
$p_c$, of the in-plane aspect ratio separating the single-domain
phase from the two-domain phase.

An important remark is in order here.
Introducing the exchange interaction, $A$, has a twofold effect: i) the
domain wall width is increased to a value $L_{dw}\approx\sqrt{A/K_1}$, where $K_1$ is
the uniaxial anisotropy constant, and ii) the domain wall energy density ({\it i.e.}, the
energy per unit wall area) is increased by a quantity of order $\sqrt{AK_1}$.
A two-domain state can be energetically favorable if the gain in
dipolar energy density coming from the two opposite domains,
of order $M_s^2 b$ (where $M_s$ is the saturation magnetization and
$2b$ is the thickness), prevails on the full domain wall energy density,
which has a dipolar contribution plus the just mentioned term of order
$\sqrt{AK_1}$. A necessary condition for the transition to occur is
that the two-domain configuration is energetically favorable when $A=0$,
because the exchange (and the anisotropy) only contribute to the cost
of the energy balance, {\it not} to the gain. If the transition does
occur when $A=0$, one can wonder whether the transition is maintained
when $A \ne 0$. One is thus led to conclude that, for comparable
values of the anisotropy and dipolar energy densities ($K_1 \approx M_s^2$),
the transition disappears because the domain wall energy density (of order $\sqrt{AK_1}$)
surely dominates on the gain (of order $M_s^2 d$). Such a no-domain rule for in-plane
configurations in ultrathin films had already been found in Refs.~\cite{EPJB,Danilo}.

However, there are systems where $K_1$ is so small that the term
$\sqrt{AK_1}$ is comparable to, or even smaller than, $M_s^2 b$:
this is just the case of Permalloy. In fact, our theoretical predictions
about the occurrence of a transition from a single-domain to a two-domain state
in rectangular Permalloy dots when increasing the in-plane aspect ratio $p$ are
confirmed by recent MFM data \cite{imre} of {\it elliptic}
patterned Permalloy elements with thickness $30$ nm,
lateral size $3~ \mu$m and $p$ varying between the values
$p = 0.13$ (elongated dot, single-domain state) and $p \lesssim
1$ (almost circular dot, two-domain state).

Finally, in Section~\ref{sec_dc} we compare the discrete
and the continuum approaches. There are some indications
that the transition from single-domain to two-domain state
may appear when the thickness of the film is comparable
to the width of the domain wall. Final considerations are
reported in Section~\ref{sec_con}.

\section{Discrete-lattice calculation of the magnetic dipolar energy
of a one-monolayer-thick rectangular element}

\begin{figure}
\includegraphics*[width=12cm,angle=0,bbllx=71pt,bblly=325pt,%
bburx=556pt,bbury=603pt]{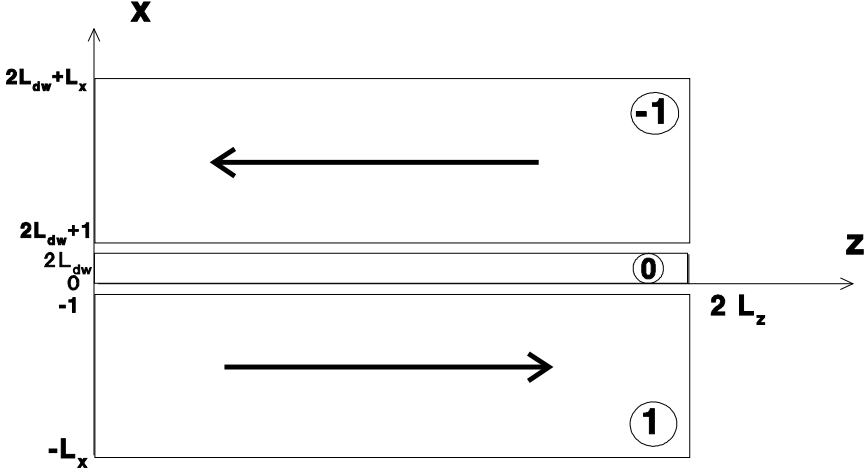} \caption{The geometry of the
rectangular element in the two-domain configuration. The bold
arrows denote the magnetization directions and $2L_{dw}$ is the
domain wall width.}
\label{stripe}
\end{figure}

Let us start considering a rectangular element which is one
monolayer thick. We want to compare the energy of the in-plane
ferromagnetic (FM) configuration, where all spins (because of
shape anisotropy) are parallel to the longer side, to the
two-domain configuration (see Fig.~\ref{stripe}), where a single
N\'eel domain wall, parallel to the longer side, is inserted in
the middle of the stripe. We have considered both a triangular and
a square lattice, with the $\hat z$ axis parallel to a row of
atoms. However, for ease of notations, we will refer to the square
lattice. Different rows are labelled with index $m$ as follows:
for $-L_x\le m\le -1$ (region `1'), the spins are parallel to the
$+\hat z$ direction, $\vec S=\hat z$; when $m$ varies between
$m=0$ and $m=2L_{dw}$ (region `0'), the spins rotate inside the
domain wall, according to the relations
$S_x(m)=\cos[m\pi/(2L_{dw})]$ and $S_y(m)=\sin[m\pi/(2L_{dw})])$;
finally, for $2L_{dw}+1\le m\le 2L_{dw}+L_x$ (region `-1'), $\vec
S=-\hat z$. If the distance between nearest neighboring spins is
taken as length unit, the size of the stripe in the $\hat z$
direction is $2L_z$ and lines parallel to $\hat z$ are distant
$1$.

The full dipolar energy of a given configuration can be written as
\begin{equation}
E_{dip}=\Omega\sum_{(ij)}{1\over r_{ij}^3}\left[
\vec S_i\cdot\vec S_j -3{(\vec S_i\cdot \vec r_{ij}) (\vec
S_j\cdot \vec r_{ij})\over r_{ij}^2}\right],
\end{equation}
where $\Omega=\mu^2/a_0^3=M_s^2 a_0^3$ ($\mu$ is the magnetic
moment per spin, $M_s$ is the magnetic moment per unit volume,
and $a_0$ is the lattice constant).

\begin{figure}
\includegraphics*[width=12cm]{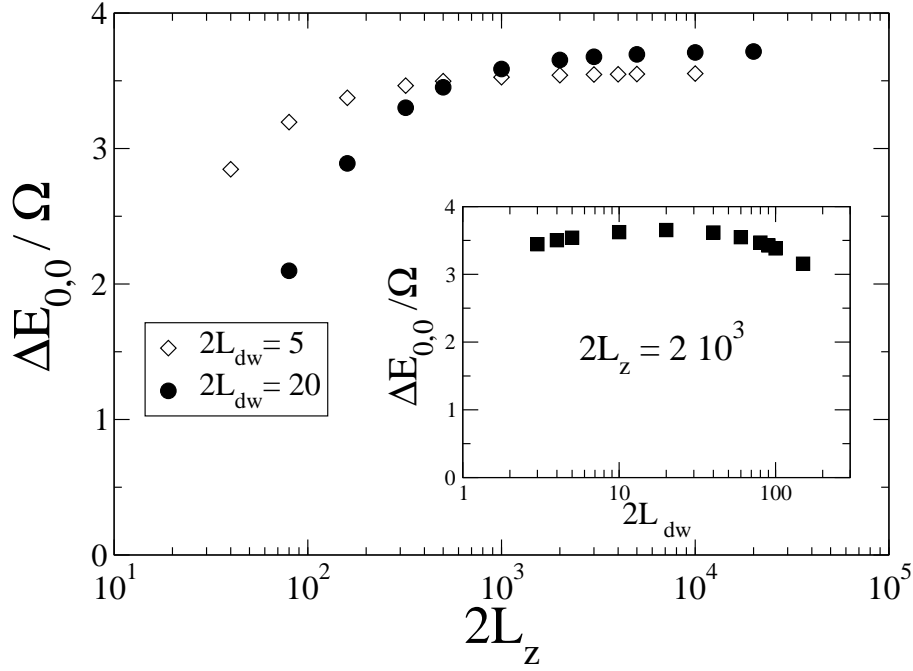}
\caption{Main: The dipolar contribution to the domain wall energy
per unit length, $\Delta E_{0,0}$, as a function of $2 L_z$, for
$2L_{dw}=5$ (circles) and $2L_{dw}=20$ (diamonds).
Inset: The same quantity, $\Delta E_{0,0}$, as a function of $L_{dw}$,
for $L_z=10^3$. All results refer to a square lattice.}
\label{fig_en_dw}
\end{figure}

Interacting spins $(ij)$ can be grouped according to the regions, `1', `0'
and `-1', they belong to, so that:
\begin{equation} E_{dip} =
E_{1,1}+E_{0,0}+E_{-1,-1}+E_{1,0}+E_{-1,0}+E_{1,-1} .
\end{equation}
The first three terms are the self-energies of the three regions;
$E_{1,0}$ and $E_{-1,0}$ are the interaction energies between each
ferromagnetic region and the domain wall region; finally,
$E_{1,-1}$ is the energy between the two ferromagnetic regions. We
are interested in the energy difference between the two-domain
state and the FM (single-domain) state. Since it will be useful to
consider energies per unit length in the $\hat z$ direction, we
define $\Delta E_{dip} = (E_{dip} - E_{dip}^{FM})/(2 L_z)$,
where $E_{dip}$ and $E_{dip}^{FM}$
are the energies of the two-domain and of the FM state, respectively.
When the symbols $E$ and $E^{FM}$ are accompanied by the subscripts `1', `0',
`-1', the energies refer to self/interaction dipolar energies of specific
regions. Since regions `1' and `-1' keep themselves ferromagnetic,
$\Delta E_{1,1}=\Delta E_{-1,-1}=0$, and
\begin{equation}
\Delta E_{dip} = \Delta
E_{0,0} + \Delta E_{1,0}+\Delta E_{-1,0}+ \Delta E_{1,-1}.
\end{equation}
As for $\Delta
E_{1,0}$ and $\Delta E_{-1,0}$ (which are equal, for symmetry
reasons), they can be neglected~\cite{review_PP,KP},
because the interaction between a FM domain and the domain wall
is averaged to zero in the limit $L_x\gg L_{dw}$.
Finally, we have
\begin{equation}
\Delta E
\simeq \Delta E_{0,0} + \Delta E_{1,-1},
\label{Delta_E}
\end{equation}
where the first term is the self-energy of the domain wall and the
second term is the interaction energy between the two FM regions.
In Fig.~\ref{fig_en_dw} (main) we show that $\Delta E_{0,0}$ goes
to a constant with increasing $L_z$, so that we can define a
dipolar domain wall energy per unit length. As for the
$L_{dw}$--dependence of $\Delta E_{0,0}$, it is shown in the inset
of the same figure. For a large range of values of $L_{dw}$, $1\ll
L_{dw}\ll L_z$, $\Delta E_{0,0}$ can be approximately taken as
constant. In conclusion, we can assume that $\Delta
E_{0,0}/\Omega$ is a constant $d_0$, with $d_0\simeq 4$.

The quantity $\Delta E_{1,-1}$ can be rewritten as follows,
\be
\Delta E_{1,-1} = (E_{1,-1} - E^{FM}_{1,-1})/(2 L_z)
= -2 E^{FM}_{1,-1}/(2 L_z),
\ee
because in the
two-domain state, the spins in the `-1' region are just reversed
with respect to the FM state. Since the two regions are $2L_{dw}$
far away, we can replace the discrete summation by an integral.
We get the result:

\begin{eqnarray}
&& E^{FM}_{1,-1}/\Omega = \cr && 2\sqrt{4L_z^2+2L_{dw}^2}
-4\sqrt{4L_z^2+(L_x+2L_{dw})^2}+2\sqrt{4L_z^2+(2 L_x+2L_{dw})^2}\cr
&-&4(L_x+2L_{dw})\ln(L_x+2L_{dw}) + 2L_{dw}\ln 2L_{dw}\cr
&-&(6L_x+8L_{dw})\ln[ \sqrt{4L_z^2+(L_x+2L_{dw})^2} -L_x-2L_{dw}]\cr
&-&2L_x\ln[ \sqrt{4L_z^2+(L_x+2L_{dw})^2} +L_x+2L_{dw} ]\cr
&+&2(2L_x+2L_{dw})\ln(2L_x+2L_{dw}) +2(L_x+2L_{dw})\ln[
\sqrt{4L_z^2+2L_{dw}^2} -2L_{dw} ] \cr &+&2L_x\ln[
\sqrt{4L_z^2+2L_{dw}^2}+2L_{dw} ]\cr &+&2(2L_x+2L_{dw})\ln[
\sqrt{4L_z^2+(2L_x+2L_{dw})^2} -2L_x-2L_{dw} ].
\end{eqnarray}
In the relevant limit $L_{dw}\ll L_z, L_x$, we get
\be
\Delta E_{1,-1}/\Omega \simeq
-4\Bigg\{ 1+ p\ln 4 -\sqrt{4+p^2} +\sqrt{1+p^2}-p
\ln\Bigg({\sqrt{4+p^2}-p\over\sqrt{1+p^2}-p}\Bigg) \Bigg\},
\label{eq_deltaE_1m1}
\ee
where $p=L_x/L_z$ is the aspect ratio of the
stripe. The above function, which gives the gain in dipolar energy
from the interaction of the two FM regions with opposite
magnetization, varies between $\Delta E_{1,-1}=0$ for $p=0$
(infinitely elongated sample) to
$\Delta E_{1,-1}\simeq -1.9\Omega$ for $p=1$
(square sample).

The two-domain state can be favored with
respect to the FM state only if $\Delta E_{dip} \simeq \Delta E_{0,0} + \Delta
E_{1,-1} < 0$. Since $\Delta E_{0,0} \approx 4\Omega$ and $\Delta
E_{1,-1}\approx -2\Omega$, the conclusion is that a two-domain state
with an {\it extended} domain wall can {\it not} appear in a one-monolayer
thick stripe, whatever the aspect ratio is.
For the triangular lattice one has $\Delta E_{0,0} \approx 5\Omega$
and $\Delta E_{1,-1}\approx -2\Omega$, so the same conclusion
can be drawn. This is not in contradiction
with what we found in Ref. \cite{PPS}, because there we assumed an
atomically {\it sharp} domain wall ($L_{dw}=0$ in the
present notations).

As stressed in the Introduction,
the reason why we are now considering $L_{dw}\gg 1$, is because
exchange interaction widens the domain wall.
It is well known \cite{aharoni_book} that in the presence of
a short range ferromagnetic interaction, $A$, and of a uniaxial
anisotropy, $K_1$ (favoring the $\pm\hat z$ directions), the
additional contribution to the domain wall energy per unit length
is $\Delta E_{ex+ani} \approx \sqrt{AK_1}>0$ and the
resulting domain wall width is $L_{dw}\approx \sqrt{A/K_1}$.
Therefore, the total energy difference (per unit length) $\Delta E_{total}$
between the two-domain and the FM single domain state is
\be
\Delta E_{total}=\Delta E_{0,0} + \Delta E_{1,-1} + \Delta E_{ex+ani}.
\ee
Since the gain in energy of a two-domain state must be provided
by dipolar energy, a positive value for $(\Delta E_{0,0} + \Delta E_{1,-1})$
can {\em not} be healed by the additional positive term $\Delta E_{ex+ani}$.
The conclusion of this Section is that a single FM monolayer does not split
into two domains if exchange interaction makes the domain wall width finite.

\section{Micromagnetic calculation of the magnetostatic energy of a rectangular prism}

\begin{figure}
\includegraphics*[width=12cm,angle=0,bbllx=34pt,bblly=331pt,%
bburx=510pt,bbury=766pt]{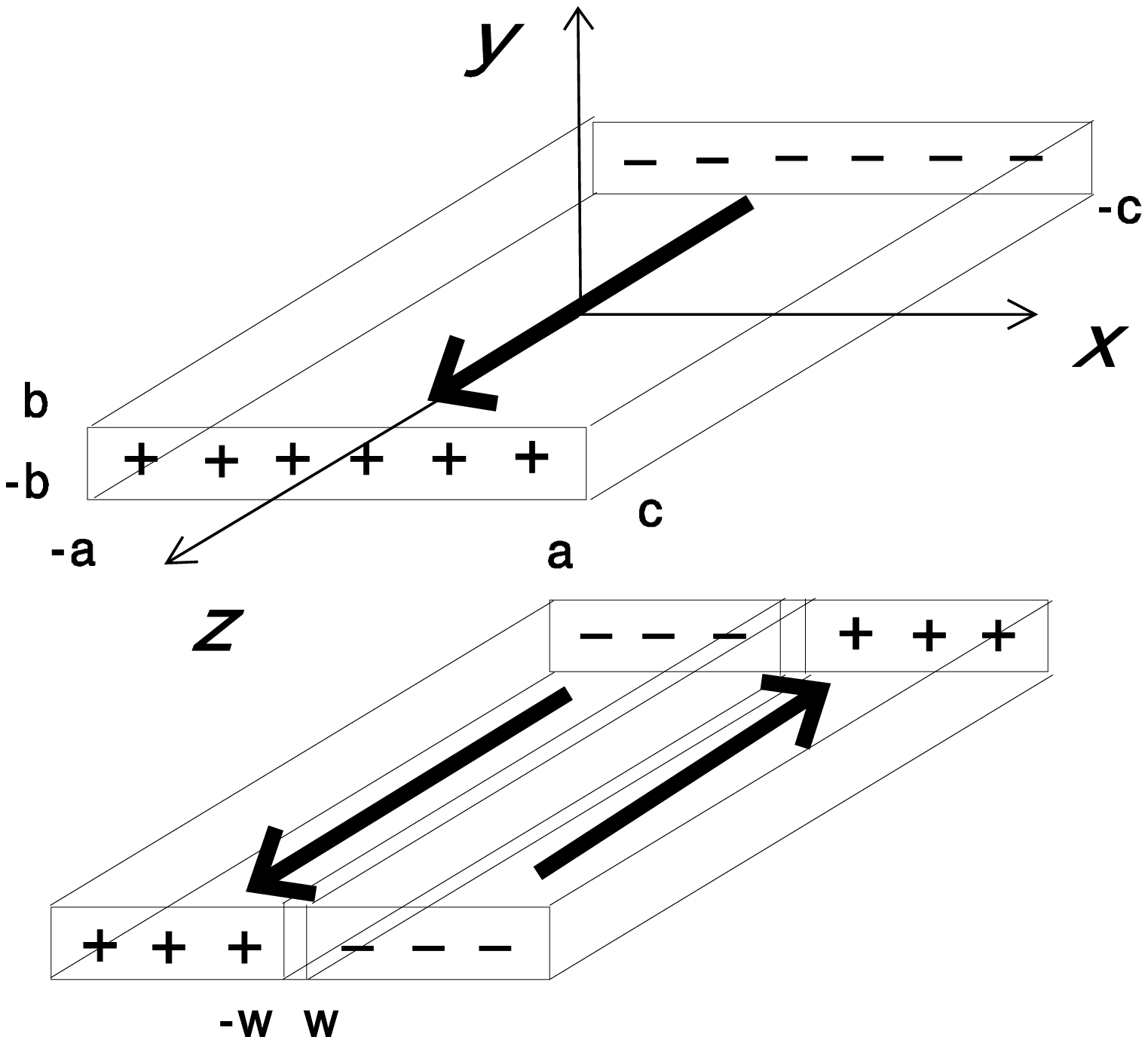}
\caption{The geometry of the
rectangular ferromagnetic prism. The bold arrows denote the
magnetization directions. Schematic representations of the surface
magnetostatic charges are displayed for the single-domain state
(top) and for the two-domain state (bottom).}
\label{prism}
\end{figure}

A few years ago, Aharoni~\cite{aharoni_jap} gave an analytic expression for the
demagnetizing factors of a uniform and homogeneous ferromagnetic
particle in the shape of a rectangular prism,
extending over the volume $-a \le x \le a$, $-b \le y \le b$ and
$-c \le z \le c$, with the origin of the Cartesian coordinate
system at the center of the prism (see Fig.~\ref{prism}, top).
With respect to the notations used in the previous Section,
we have $a=a_0 L_x$, $b=a_0 L_y$, $c=a_0 L_z$, and
$w=a_0 L_{dw}$ (see footnote\footnote{%
Since $L_{dw}\ll L_x$, the definitions $a=a_0 L_x$ and
$a=a_0(L_x+L_{dw})$ are equivalent.}).
Here we briefly sketch Aharoni's calculation~\cite{aharoni_jap}
for clarity's sake.

When the prism is saturated along $+z$, surface charges $\pm M_s$ are
created on the faces $z=\pm c$,
where $M_s$ is the saturation magnetization.
The potential due to the surface charges is (in cgs units)~\cite{aharoni_book}
\begin{equation}
U({\bf r})= \int_S { { {\bf n}\cdot{\bf
M}({\bf r}^{\prime}) }\over {\vert {\bf r}-{\bf r}^{\prime} \vert}
}~ dS^{\prime}.
\end{equation}
The density of the
surface charges is $({\bf n}\cdot {\bf M})$ where ${\bf M}$ is the
magnetization and ${\bf n}$, the unit normal to the surface $S$ of
the ferromagnetic body, is taken to be positive in the outward
direction. The magnetic field generated by the surface charge
distribution is ${\bf H}=-{\bf \nabla} U$ and the magnetostatic
self-energy is given by
\begin{equation}
\label{energy} E_M=-{1\over 2} \int_V {\bf M} \cdot {\bf H}~
dV^{\prime},
\end{equation}
where the integration is over the volume $V$ of the ferromagnetic
body. As a general rule \cite{aharoni_book,aharoni_ieee}, it is
not advisable to evaluate first the potential $U$ for a specific
case, then obtain the magnetostatic field ${\bf H}=-\nabla U$, and
finally substitute ${\bf H}$ in Eq.~(\ref{energy}). It is better
to write (\ref{energy}) with all its integrals and then consider
which to perform first. In other words, the order in which
integrations are carried can considerably change the amount of
algebra. The magnetostatic self-energy of the uniform
rectangular prism depicted in Fig.~\ref{prism} (top) is
\begin{eqnarray}
&&E_M^{\uparrow \uparrow}=
{1\over 2} M_s^2 \int_{-c}^{+c} dz {{\partial}\over
{\partial z}}\int_{-b}^{b} dy\int_{-b}^b d{\eta} \int_{-a}^{a} dx
\int_{-a}^{a} d{\xi} \cr && \bigg[ ~{1\over
{\sqrt{(x-{\xi})^2+(y-{\eta})^2+(z-c)^2}}} -{1\over
{\sqrt{(x-{\xi})^2+(y-{\eta})^2+(z+c)^2}}}~ \bigg]\nonumber.
\end{eqnarray}
By carrying out analytically the four integrations over $x$, $y$,
$\xi$, $\eta$ (see Appendix A for details) one obtains
\begin{equation}
E_M^{\uparrow \uparrow}={1\over 2}
M_s^2~V N_z ,
\end{equation}
where the explicit expression for $N_z$, the demagnetizing factor
of the general rectangular prism of volume $V=8 a b c$, is given
in Ref.~\cite{aharoni_jap}. In general one has $N_x+N_y+N_z=4\pi$;
for a cubic sample ($a=b=c$), $N_x=N_y=N_z={{4\pi}\over 3}$.

When the
rectangular prism is divided into two domains with opposite
magnetizations directed along $\pm z$ (see Fig.~\ref{prism}, bottom), the
contribution of surface charges to the magnetostatic energy is
\begin{eqnarray}
&&E_M^{\downarrow \uparrow}={1\over 2}
M_s^2 \Bigg\{ \int_{-c}^{+c} dz {{\partial}\over {\partial z}}
\int_{-b}^{b} dy\int_{-b}^b d{\eta} \int_{-a}^{a} dx
\int_{-a}^{-w} d{\xi} \cr && \bigg[ {1\over
{\sqrt{(x-{\xi})^2+(y-{\eta})^2+(z-c)^2}}} -{1\over
{\sqrt{(x-{\xi})^2+(y-{\eta})^2+(z+c)^2}}}\bigg] \cr &&-
\int_{-c}^{+c} dz {{\partial}\over {\partial z}} \int_{-b}^{b}
dy\int_{-b}^b d{\eta} \int_{-a}^{a} dx \int_{w}^{a} d{\xi} \cr &&
\bigg[ {1\over {\sqrt{(x-{\xi})^2+(y-{\eta})^2+(z-c)^2}}} -{1\over
{\sqrt{(x-{\xi})^2+(y-{\eta})^2+(z+c)^2}}} \bigg] \Bigg\}.
\nonumber
\end{eqnarray}

Taking into account the symmetry properties of the multiple integrals (see
Appendix A), one obtains $E_M^{\downarrow \uparrow}= 0$.
Thus, the difference $\Delta E_M=
E_M^{\downarrow \uparrow}-E_M^{\uparrow \uparrow}$ between the
surface magnetostatic energy of the prism in the two-domain
configuration and the magnetostatic energy of the same prism in
the single-domain configuration is given by
\begin{equation}
\label{deltaE_magnetostatic} \Delta E_M= E_M^{\downarrow
\uparrow}-E_M^{\uparrow\uparrow}= -
{1\over 2} M_s^2 V N_z,
\end{equation}
{\it i.e.}, it is just the opposite of the magnetostatic
self-energy of the single-domain prism.
The explicit expression of $\Delta E_M$,
as a function of the reduced thickness $t=b/c$
and of the in-plane aspect ratio $p=a/c$ of
the rectangular prism, turns out to be
\begin{eqnarray}
\label{explicit} &&\Delta E_M= E_M^{\downarrow \uparrow}
-E_M^{\uparrow \uparrow}\approx -{{\gamma_B}\over {4\pi}}{1\over
2} M_s^2 V \Bigg[ 8\arctan \bigg( {{{pt}\over {\sqrt{1+p^2+t^2}}}}
\bigg)\cr &+&{{2(t^2-1)}\over {t}} \ln \bigg(
{{{\sqrt{1+p^2+t^2}-p}\over {\sqrt{1+p^2+t^2}+p}}} \bigg)
+{{2(p^2-1)}\over {p}} \ln \bigg( {{{\sqrt{1+p^2+t^2}-t}\over
{\sqrt{1+p^2+t^2}+t}}} \bigg) \cr &-& 2t \ln \bigg(
{{{\sqrt{p^2+t^2}-p}\over {\sqrt{p^2+t^2}+p}}} \bigg) - 2p \ln
\bigg( {{{\sqrt{p^2+t^2}-t}\over {\sqrt{p^2+t^2}+t}}} \bigg) \cr
&+& {{2}\over {p}} \ln \bigg( {{{\sqrt{1+t^2}-t}\over
{\sqrt{1+t^2}+t}}} \bigg) + {{2}\over {t}} \ln \bigg(
{{{\sqrt{1+p^2}-p}\over {\sqrt{1+p^2}+p}}} \bigg)  \cr &+&
{{4(p^3+t^3-2)}\over {3 pt}}+ {{4(p^2+t^2-2)}\over {3 pt
}}\sqrt{1+p^2+t^2}+{{4}\over {pt}} (\sqrt{1+p^2}+\sqrt{1+t^2})\cr
&-&{4\over {3pt}}\big( (p^2+t^2)^{3\over 2}+(1+p^2)^{3\over
2}+(1+t^2)^{3\over 2}\big) \Bigg].
\end{eqnarray}

Clearly, the quantity $\Delta E_M$ (see Eq.~\ref{deltaE_magnetostatic})
is always negative: as for the surface contribution
to the magnetostatic self-energy, the system would prefer to
divide in two domains.
In the limit of vanishing thickness, the magnetostatic energy difference
$\Delta E_M$ approaches 0, since for $t \to 0$ one has
\begin{eqnarray}
&N_z&  \approx -4t \ln t \cr &+& t \Bigg\{ 4 \ln (2p) -{4\over
p}+2+{4\over p}\sqrt{1+p^2} +2 \ln [1+2p^2\sqrt{1+p^2}]\Bigg\} +
O(t^2).
\label{eq_Nz}
\end{eqnarray}
The vanishing of $\Delta E_M$ could be expected on general grounds
because the number of surface charges tends to 0 as the height of the prism shrinks
\cite{squeeze}.

A limit to the splitting of the sample in two domains is posed by
the full cost of the domain wall energy, which involves other energy terms:
exchange interaction and anisotropy. The first term (exchange)
is short ranged and favours parallel alignment of neighboring spins:
\begin{equation}
\label{exchange} {\it E}_{ex}={A\over {M_s^2}} \int_V ({\bf
\nabla} \cdot {\bf M})^2~dV^{\prime}.
\end{equation}

The anisotropy term, assumed to favor the alignment of the magnetization along
the $z$ axis, has the form
\begin{equation}
\label{anisotropy}
 {\it E}_{ani}=-{{K_1}\over {M_s^2}} \int_V (M^z)^2~dV^{\prime}.
\end{equation}

The magnetostatic contribution to the N\'eel domain wall energy
is determined assuming a one-dimensional model of
the wall~\cite{aharoni_book},
{\it i.e.}, the magnetization within the wall ($-w \le x
\le w$) is a function of $x$ only.
The components of the unit magnetization are assumed to be
\begin{equation}
m_x(x)={{w^2}\over {w^2+x^2}},~~ m_y(x)=0,~~
m_z(x)={{x\sqrt{2w^2+x^2}}\over {w^2+x^2}},
\end{equation}
where $w$ determines
the wall width. At the ends of the wall, where the domains begin,
one has $m_z(\pm \infty)=\pm 1$.  In a thin film, the
magnetostatic energy of the volume charges within a N\'eel wall
which occupies the region $-w \le x \le w$
can be approximated by
\cite{aharoni_book}
\begin{eqnarray}
{\it E}_m &=& 4\pi M_s^2 b \int_{-\infty}^{\infty}dx [m_x(x)]^2
\cr
&& - 2M_s^2
\int_0^{\infty}dt{{1-e^{-2bt}}\over {t}}\int_{-\infty}^{\infty} dx
\int_{-\infty}^{\infty}dx^{\prime}~ \cos[(x-x^{\prime})t]~ m_x(x)
m_x(x^{\prime}),
\end{eqnarray}
provided that the domain wall width is much smaller than the
lateral dimensions of the sample, $w \ll a \le c$.
In the same approximation, the exchange and the uniaxial anisotropy
contributions to the domain wall energy are, respectively
\begin{eqnarray}
{\it E}_{ex}&=&A \int_{-b}^bdy \int_{-\infty}^{\infty}dx \Bigg[
\bigg( {{dm_x}\over {dx}}\bigg)^2+\bigg( {{dm_z}\over
{dx}}\bigg)^2\Bigg], \cr {\it E}_{ani}&=& K_1\int_{-b}^b dy
\int_{-\infty}^{\infty}dx~  m_x^2.
\end{eqnarray}
One finally obtains the
following expressions for the magnetostatic, the exchange and the
anisotropy contributions
\begin{eqnarray}
\label{gamma} \gamma_m&=&{{{\it E}_m}\over {4bc}} = \pi^2 M_s^2 w
\Bigg[ 1 -{w\over b} \ln \Big(1+{b\over w}\Big)\Bigg], \\
\gamma_{ex}&=&{{{\it E}_{ex}}\over {4bc}}={{2\pi A}\over
{w}}(\sqrt{2}-1),\\
\gamma_{ani}&=&{{{\it E}_{u}}\over {4bc}}={{\pi w }\over {2}}K_1,
\label{gamma2}
\end{eqnarray}
to the total N\'eel wall energy $\gamma_N=\gamma_m+\gamma_{ex}+\gamma_{ani}$
per unit wall area ($4bc$). The actual domain wall width is determined
by minimizing $\gamma_N$ with respect to $w$, thus
leading to the transcendental equation~\cite{aharoni_book}
\begin{equation}
\label{transcendental} {{2A(\sqrt{2}-1)}\over {\pi M_s^2 w^2}}
-{{K_1}\over {2\pi M_s^2}} - 1+ {{2w}\over {b}} \ln \Big(1+{b\over
w}\Big) -{{w}\over {w+b}}=0 .
\end{equation}

We have solved Eq.~(\ref{transcendental}) using the material
parameters of Permalloy ({\it i.e.}, exchange constant
$A=10^{-6}$ erg/cm, uniaxial anisotropy constant $K_1=10^3$
erg/cm$^3$, saturation magnetization $M_s=800$ emu).

\begin{figure}
\includegraphics*[width=12cm,angle=0,bbllx=82pt,bblly=216pt,%
bburx=473pt,bbury=710pt]{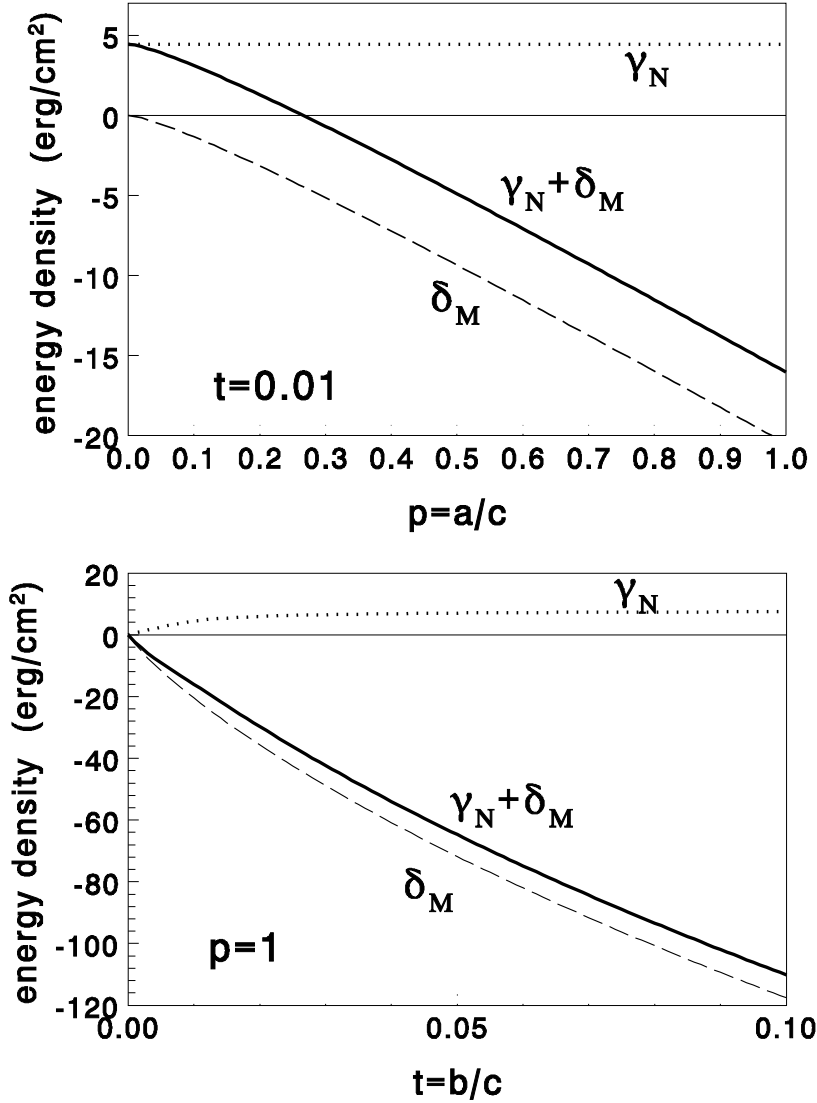}
\caption{Top: energy per unit wall area ($4bc$) of a
rectangular ferromagnetic Permalloy prism versus the in-plane aspect ratio
$p=a/c$ for fixed thickness.
The dotted line is the cost in energy density, $\gamma_N$,
of a N\'eel wall separating two opposite domains
(see Eqs.~(\ref{gamma}-\ref{gamma2})), the dashed line
is the gain in magnetostatic energy density,
$\delta_M={{\Delta E_M}\over {4bc}}$
(see Eq.~(\ref{explicit})), and the full line is the total energy density
$\gamma_N+\delta_M$. The single-domain state is
energetically favored when $p<p_c=0.27$, while the two-domain state is
preferred for $p_c<p \lesssim 1$.
Bottom: energy per unit wall area ($4bc$) of a  square ($p=a/c=1$)
ferromagnetic Permalloy prism versus the reduced thickness $t=b/c$.
In this case, the two-domain state is always preferred.
}
\label{vspvst}
\end{figure}

In Fig.~\ref{vspvst} (top) we plot the energy density ({\it i.e.},
the energy per unit wall area $4bc$) versus the in-plane aspect ratio
$p=a/c$ of a rectangular prism (larger side $2c=3\times 10^{-4}$ cm),
for fixed dot
thickness ($2b=3\times 10^{-6}$ cm). Using the Permalloy material parameters,
the calculated energy density turns out to be positive for $p<p_c$
and negative for
$p>p_c$, with $p_c=0.27$. Therefore, an elongated sample
($p<p_c$) prefers to assume a single-domain configuration since
the energy density cost of the N\'eel wall, $\gamma_N$, exceeds
the gain in surface magnetostatic energy
density, $\delta_M={{\Delta E_M}\over {4bc}}$, which is obtained
splitting the sample into two domains. Upon increasing the
in-plane aspect ratio from the critical value $p_c$ to $p=1$
(limit of a square dot), there is an increasing gain in the
surface magnetostatic energy, and the element prefers to assume a
two-domain configuration. In Fig.~\ref{vspvst} (bottom) the energy density
of a square Permalloy prism (side $2a=2c=3~10^{-4}$ cm) is plotted as a function
of the reduced thickness $t=b/c$. This means that for a square
sample the two-domain state is always preferred, whatever is the thickness.

These theoretical results for rectangular Permalloy prisms appear
to account for recent MFM data \cite{imre} in a
series of {\it elliptic} patterned Permalloy dots with thickness
$2b=30$ nm. In these samples, one axis of the elliptic elements was kept the
same ($2c = 3~ \mu$m) and the other was varied, so that a series of dots with
in-plane aspect ratio $p=a/c$ ranging between $p=0.13$ and $p=1$ was obtained.
Moreover, the interdot distance was kept sufficiently large so as
to consider the dots as non interacting. The MFM images clearly
showed that elongated dots with $p \le 0.22$ are in a single-domain
configuration, while dots with $p \ge 0.28$ are in a two-domain
configuration, in fair agreement with our theoretical result of a critical
aspect ratio $p_c=0.27$. For circular dots, a vortex configuration turned out
to be preferred \cite{imre}.

\begin{figure}
\includegraphics*[width=12cm]{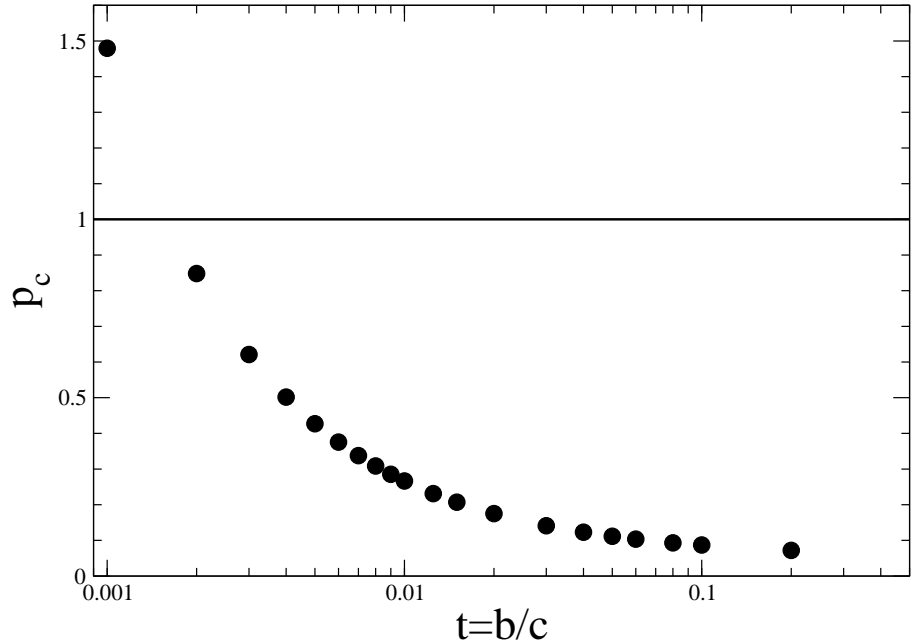}
\caption{The critical in-plane aspect ratio $p_c$ as a function of
the reduced thickness $t$, for Permalloy films. A value of
$p_c$ larger than 1 (as it is for very small thicknesses)
means that the system always stays in a single domain state:
this is an effect of the exchange$+$anisotropy cost of the domain wall.}
\label{fig_pc}
\end{figure}

The shape of the sample is fully determined by the in-plane
aspect ratio $p=a/c$ and by the reduced thickness $t=b/c$.
In Fig.~\ref{fig_pc} we plot the critical aspect ratio $p_c$
as a function of $t$, for Permalloy samples. For not so small thicknesses
($t\sim 0.1$), the sample is always in a two-domain state except for
very elongated systems ($p<0.1$). With decreasing the thickness,
the value of $p_c$ increases and for $t<0.0015$, $p_c>1$, which
means that the system always stays in the single-domain state.
This is due to the additional exchange$+$anisotropy cost of the
domain wall, because, as it appears from Eq.~(\ref{deltaE_magnetostatic})
and Eq.~(\ref{eq_Nz}) (see also the next Section),
the balance of dipolar energy is always in favor of the two-domain
state for a square prism ($p=1$).

\section{Changing the film thickness: discrete vs continuum approaches}
\label{sec_dc}

In Section 2 we have shown that a one ML thick magnetic particle
does not undergo a splitting from a single-domain to a two-domain
state with passing from an elongated to a square sample.
In Section 3, within a continuum approach, we have shown
that such splitting does exist for a film. Therefore, we should be
led to conclude that this transition appears at some critical
thickness. In the following we are going to discuss this problem.

First, we observe that the key quantities to be compared are $(E_{1,-1}-
E_{1,-1}^{FM})$, $(E_{0,0}-E_{0,0}^{FM})$ in the discrete
approach, and $(E_M^{\downarrow
\uparrow}-E_M^{\uparrow\uparrow})$, $E_m$ in the continuum approach.
The quantities $(E_{1,-1}- E_{1,-1}^{FM})$ and $(E_M^{\downarrow
\uparrow}-E_M^{\uparrow\uparrow})$ are negative and mean the
gain in dipolar energy when two domains of opposite
magnetization are formed.
The other quantities, $(E_{0,0}-E_{0,0}^{FM})$ and $E_m$,
are positive and mean the dipolar cost in forming a domain wall.
A square-shaped particle can possibly split in two domains only if the
dipolar gain is larger than the dipolar cost.
In the following we are comparing
discrete and continuum expressions of the dipolar gain and
of the dipolar cost for a square particle.\footnote{An elongated
film ($p\ll 1$) is surely in a single domain state. A transition
to a two-domain state occurs if and only if a square particle is
in a two-domain state.}

\subsection{The dipolar gain}

According to Eq.~(\ref{deltaE_magnetostatic}),
$\Delta E_M= E_M^{\downarrow \uparrow}-E_M^{\uparrow\uparrow}
=-{1\over 2}M_s^2 V N_z$. The quantity $N_z$ is a complicated
expression, but we can confine here to $p=1$ (square prism) and
$t\to 0$ (vanishing thickness). In these limits, see Eq.~(\ref{eq_Nz}),
\be
\Delta E_M=-{1\over 2}M_s^2 V N_z \simeq -{1\over 2}M_s^2 V
( -4t\ln t +\bar c t),
\ee
with $\bar c \simeq 2.9$.
In the ML limit, see  Eq.~(\ref{eq_deltaE_1m1}), we had
\be
\label{MLlimit}
E_{1,-1}-E_{1,-1}^{FM} \simeq - 1.9 \Omega (2L_z).
\ee
It is worth noting that for both approaches, in
the limit of a domain wall width much smaller than the in-plane size
of the particle, the dipolar energy gain does not depend on the
domain wall width.

If we consider the case of domain wall width much larger than
film thickness, $L_{dw} \gg L_y$, the result (\ref{MLlimit}) for the ML
can be extended to finite film thickness simply multiplying it
by $(2L_y)^2$, because there are $(2L_y)$ stripes magnetized in
the $\hat z$ direction which interact with the $(2L_y)$ stripes
magnetized in the opposite direction, so that
\be
E_{1,-1}-E_{1,-1}^{FM} \simeq - 1.9 \Omega (2L_z) (2L_y)^2
\simeq -{1\over 2} M_s^2 V (3.8 t).
\ee
Therefore, the discrete approach is directly comparable with the
continuum one: the former gives $N_z\approx 3.8 t$ and the
latter $N_z\approx -4t\ln t +2.9 t$.

\subsection{The dipolar cost}

According to Eq.~(\ref{gamma}), the dipolar cost of the N\'eel wall is
\be
E_m = 4bc~
\pi^2 M_s^2 w
\Bigg[ 1 -{w\over b} \ln \Big(1+{b\over w}\Big)\Bigg],
\ee
where $2w$ is the domain-wall width and $2b$ is the thickness.
In the relevant limit $w\gg b$, we get
\be
E_m \approx {\pi^2\over 4} M_s^2 V t.
\ee
In the discrete approach for the ML, see Eq.~(\ref{Delta_E})
and Fig.~2, we have the corresponding quantity
\be
E_{0,0} - E_{0,0}^{FM} \simeq \Omega d_0 (2L_z),
\ee
with $d_0\approx 5$.
Therefore, for both approaches, the dipolar domain wall energy does not
depend on $w$ in the limit $w\ll a$ (or, equivalently, $L_{dw}\ll L_x$).

Extending the discrete approach to a finite thickness, we have to
distinguish between interacting spins belonging
to the same layer and to different layers.
There are $(2L_y)$ planes and $(2L_y)(2L_y-1)/2$
distinct pairs of planes; within the
usual approximation $L_{dw}\gg L_y$,
the interaction between different planes within the domain wall does not
depend on their distance, so that
\be
E_{0,0} - E_{0,0}^{FM} \simeq \Omega (2L_z)
\left[ d_0 (2L_y) + d_1 {(2L_y)(2L_y-1)\over 2}\right]
\approx {d_1\over 2} M_s^2 V t,
\ee
where we have retained only the leading term.

In the above expressions, $d_0\approx 5$,
while $d_1$ represents the interaction energy
between different layers within the domain wall.
We have numerically checked that $d_1>5$, therefore
the dipolar energy cost $(d_1/2) M_s^2 V t$
always dominates on the dipolar gain $(-1.9 M_s^2 V t)$.
On the other hand, in the continuum picture we have a
dipolar gain $\Delta E_M = -{1\over 2} M_s^2 V
(-4t\ln t + 2.9 t)$ which dominates on the dipolar cost
$E_m = (\pi^2/4) M_s^2 V t$ for any thickness $t$.

To sum up, the continuum approach suggests that a square
particle is splitted in two domains for any thickness,
while the discrete approach supports the opposite conclusion.
However, {\it extrapolation of the discrete results from
the ML to finite thickness is valid only within the limit
of thickness much smaller than the domain wall width}.
We are therefore led to conclude that a critical thickness
$L_y^* \approx L_{dw}$ should exist, such that the transition
between single-domain to two-domain state appears at
thicknesses $L_y > L_y^*$ only.

\section{Conclusions}
\label{sec_con}

In Ref.~\cite{PPS} we had considered a single monolayer
with planar spins interacting only via dipolar interaction.
For a triangular lattice, it was possible to observe a
transition from a single-domain to a two-domain to a macrovortex
state, when passing from elongated to square (or circular) samples.
This phenomenology did not apply to a square lattice,
because in that case spins do not like ferromagnetic configurations.
The domain wall in the two-domain state was atomically sharp.

In this paper we have studied in detail the transition from
a single-domain to a two-domain configuration for a film
of general thickness and in the presence of exchange interaction
and single ion anisotropy. The presence of the exchange interaction
has a twofold effect: it widens the domain wall and, together with the
anisotropy, it increases the energetic cost of the domain wall.
The first effect (a larger domain wall) is enough to suppress
the transition from single domain to two-domain state in a monolayer
and in ultrathin films. These results have been obtained within
a discrete approach.

For thicker films a continuum, micromagnetic approach is more
appropriate. According to it, the energetic balance of the
sole dipolar interaction is such that the above transition
always occurs on passing from elongated to square samples.
When exchange and anisotropy are correctly taken into account
in the full energy balance, it is straightforward to realize
that the transition may only occur for systems where the
anisotropy $K_1$ is vanishing small, because the exchange $A$ is usually
so large that the additional cost in domain wall energy density
$\sqrt{AK_1}$ can not be compensated by the gain in dipolar energy density,
of order $M_s^2 b$.
A material satisfying the condition $\sqrt{A K_1} \ll M_s^2 b$ is Permalloy.
For rectangular prisms made of this material, we have therefore determined
the critical in-plane aspect ratio $p_c$ as a function of the prism thickness.
Fair agreement with existing experimental data on Permalloy particles \cite{imre}
was obtained.

A final comment concerns the discrete and the continuum approaches.
The former has been applied to ultrathin films and has shown that
increasing the domain-wall width suppresses the transition.
In contrast, the transition is obtained using the latter approach,
valid for thicker films.
Therefore, these results suggest the existence of some critical
thickness $d^*$ above which the transition appears. The discrete formalism
suggests that this thickness $d^*$ should be of order of the domain wall
width $w$, but the continuum formalism does not seem to indicate
that something ``critical" should occur when $b \approx w$.
This is the sole issue we have not been able to clarify in full.

\ack Work supported by Italian MIUR and CNR (FISR Project
``Nanotecnologie per dispositivi di memoria ad altissima
densit\`a" and FIRB Project ``Microsistemi basati su materiali
magnetici innovativi strutturati su scala nanoscopica").

\appendix
\section{Analytical calculation of the multiple integrals}

The magnetostatic self-energy of the distributions of surface
charges depicted in Fig.~\ref{prism} can be expressed in terms of the
multiple integrals
\begin{eqnarray}
&&I^+(w,a,b,k)=\int_{-b}^{b} dy~\int_{-b}^b d{\eta} \int_{-a}^{+a}
dx~\int_{w}^{a} d{\xi}~  {1\over
{\sqrt{(x-{\xi})^2+(y-{\eta})^2+k^2}}}\cr
&&I^-(w,a,b,k)=\int_{-b}^{b} dy~\int_{-b}^b d{\eta} \int_{-a}^{+a}
dx~\int_{-a}^{-w} d{\xi}~  {1\over
{\sqrt{(x-{\xi})^2+(y-{\eta})^2+k^2}}}
\end{eqnarray}

By symmetry, it immediately turns out that
$I^+(w,a,b,k)=I^-(w,a,b,k)=I(w,a,b,k)$ and that
$I(w,a,b,-k)=I(w,a,b,k)$. The explicit expression of $I(w,a,b,k)$
is
\begin{eqnarray}
\label{I} I(a,b,k)&=& {2\over 3} (k^2)^{3\over 2}-{1\over
3}(4a^2+4b^2+k^2)^{3\over 2} +{1\over 3} (4a^2+k^2)^{3\over
2}+{1\over 3} (4b^2+k^2)^{3\over 2} \cr &-& {1\over 3} \Bigg[
((w+a)^2+k^2)^{3\over 2} -((w+a)^2+4b^2+k^2)^{3\over 2}\cr
&-&((w-a)^2+k^2)^{3\over 2} +((w-a)^2+4b^2+k^2)^{3\over2} \Bigg]
\cr &+&k^2\sqrt{4a^2+4b^2+k^2} -k^2
\sqrt{4a^2+k^2}-k^2\sqrt{4b^2+k^2} \cr &+&k^2\Bigg[
\sqrt{(w+a)^2+k^2}-\sqrt{(w+a)^2+4b^2+k^2} \cr &-&
\sqrt{(w-a)^2+k^2}+\sqrt{(w-a)^2+4b^2+k^2} \Bigg] \cr &-& 8abk
\arctan\Bigg({{4ab}\over {k\sqrt{4a^2+4b^2+k^2}}}\Bigg) \cr
&+&4bk\Bigg[(w+a) \arctan\Bigg( {{2(w+a)b}\over
{k\sqrt{(w+a)^2+4b^2+k^2}}} \Bigg) \cr &-&(w-a) \arctan\bigg(
{{2(w-a)b}\over {k\sqrt{(w-a)^2+4b^2+k^2}}} \Bigg) \Bigg] \cr &+&
bk^2 \ln\Bigg({{\sqrt{4b^2+k^2}+2b}\over
{\sqrt{4b^2+k^2}-2b}}\Bigg) \cr &-& bk^2 \ln\Bigg(
{{\sqrt{4a^2+4b^2+k^2}+2b}\over {\sqrt{4a^2+4b^2+k^2}-2b}}\Bigg)
\cr &+& 4a^2b \ln\Bigg({{\sqrt{4a^2+4b^2+k^2}+2b}\over
{\sqrt{4a^2+4b^2+k^2}-2b}}\Bigg) \cr &+& 2ak^2
\ln\Bigg(\sqrt{4a^2+k^2}+2a\Bigg) \cr &+&k^2\Bigg[ -(w+a)
\ln\Bigg(\sqrt{(w+a)^2+k^2}-(w+a)\Bigg)\cr &+&(w-a)
\ln\Bigg(\sqrt{(w-a)^2+k^2}-(w-a)\Bigg)\Bigg] \cr &-& 2ak^2
\ln\Bigg(\sqrt{4a^2+4b^2+k^2}+2a\Bigg) \cr &+&k^2\Bigg[(w+a) \ln
\Bigg({\sqrt{(w+a)^2+4b^2+k^2}-(w+a)}\Bigg)\cr &-&(w-a)
\ln\Bigg({\sqrt{(w-a)^2+4b^2+k^2}-(w-a)}\Bigg)\Bigg] \cr &-& 8ab^2
\ln\Bigg(\sqrt{4a^2+4b^2+k^2}-2a\Bigg) \cr &+&4b^2 \Bigg[(w+a) \ln
\Bigg({\sqrt{(w+a)^2+4b^2+k^2}-(w+a)}\Bigg)\cr &-&(w-a)
\ln\Bigg({\sqrt{(w-a)^2+4b^2+k^2}-(w-a)}\Bigg)\Bigg] \cr
&+&b\Bigg[(k^2-(w+a)^2)\ln\Bigg({{\sqrt{(w+a)^2+4b^2+k^2}+2b}\over
{\sqrt{(w+a)^2+4b^2+k^2}-2b}} \Bigg)\cr &-&(k^2-(w-a)^2)\ln\Bigg(
{{\sqrt{(w-a)^2+4b^2+k^2}+2b}\over {\sqrt{(w-a)^2+4b^2+k^2}-2b}}
\Bigg)\Bigg]\cr &-&k^2 \Bigg[
(w+a)\ln\Bigg({{\sqrt{(w+a)^2+k^2}+(w+a)}\over
{\sqrt{(w+a)^2+k^2}-(w+a)}} \Bigg)\cr &+&(w-a)\ln\Bigg(
{{\sqrt{(w-a)^2+k^2}-(w-a)}\over {\sqrt{(w-a)^2+k^2}+(w-a)}}
\Bigg) \Bigg]\cr &+&k^2\Bigg[(w+a)\ln \Bigg(
{{\sqrt{(w+a)^2+4b^2+k^2}+(w+a)}\over
{\sqrt{(w+a)^2+4b^2+k^2}-(w+a)}}\Bigg) \cr &+& (w-a)\ln\Bigg(
{{\sqrt{(w-a)^2+4b^2+k^2}-(w-a)}\over
{\sqrt{(w-a)^2+4b^2+k^2}+(w-a)}} \Bigg) \Bigg]
\end{eqnarray}

Taking the limit $w \to 0$ in the terms in square parentheses on
the r.h.s., one obtains
\begin{eqnarray}
\label{I0} I(0,a,b,k)&=& {2\over 3} (k^2)^{3\over 2}-{1\over 3}
(4a^2+4b^2+k^2)^{3\over 2} \cr  &+&{1\over 3} (4a^2+k^2)^{3\over
2} +{1\over 3} (4b^2+k^2)^{3\over 2} \cr
&+&k^2\sqrt{4a^2+4b^2+k^2}\cr &-& k^2
\sqrt{4a^2+k^2}-k^2\sqrt{4b^2+k^2}\cr &-& 8abk \arctan\Bigg(
{{4ab}\over {k\sqrt{4a^2+4b^2+k^2}}}\Bigg) \cr  &+& bk^2 \ln\Bigg(
{{\sqrt{4b^2+k^2}+2b}\over {\sqrt{4b^2+k^2}-2b}}\Bigg)\cr  &-&
bk^2 \ln\Bigg( {{\sqrt{4a^2+4b^2+k^2}+2b}\over
{\sqrt{4a^2+4b^2+k^2}-2b}}\Bigg)\cr  &+& 4a^2b \ln\Bigg(
{{\sqrt{4a^2+4b^2+k^2}+2b}\over {\sqrt{4a^2+4b^2+k^2}-2b}}\Bigg)
\cr &+& ak^2 \ln\Bigg( {{\sqrt{4a^2+k^2}+2a}\over
{\sqrt{4a^2+k^2}-2a}}\Bigg) \cr &-& ak^2 \ln\Bigg(
{{\sqrt{4a^2+4b^2+k^2}+2a}\over
{\sqrt{4a^2+4b^2+k^2}-2a}}\Bigg)\cr &+& 4ab^2 \ln\Bigg(
{{\sqrt{4a^2+4b^2+k^2}+2a}\over {\sqrt{4a^2+4b^2+k^2}-2a}}\Bigg)
\end{eqnarray}

Dropping the $a,b$ arguments for brevity's sake, and taking into
account the symmetry properties of the multiple integrals, the
magnetostatic self-energy for the one-domain configuration (see
Fig.~\ref{prism}, top) is expressed as
\begin{eqnarray}
\label{prima} &&E_M^{\uparrow \uparrow}={{\gamma_B}\over
{4\pi}}~{1\over 2}~ M_s^2 \Big[ I^-(0,0)-I^-(0,-2c)
+I^+(0,0)-I^+(0,-2c)\cr &-&I^-(0,2c)+I^-(0,0)
-I^+(0,2c)+I^+(0,0)\Big]\cr &=&{{\gamma_B}\over {4\pi}}~
M_s^2~2\Big[ I(0,0)-I(0,2c)\Big]
\end{eqnarray}

Upon substituting the explicit expression of the multiple integral
(\ref{I0}) in the equation (\ref{prima}), it is immediate to
recover Aharoni's formula for the demagnetizing factor of the
uniform and homogeneous ferromagnetic rectangular prism (Eq.~1 of
Aharoni's paper) \cite{aharoni_jap}.

For the two-domain configuration (see Fig.~\ref{prism}, bottom), the
contribution of the surface charges to the magnetostatic energy is
found to be
\begin{eqnarray}
&&E_M^{\downarrow \uparrow}={{\gamma_B}\over {4\pi}}~{1\over 2}~
M_s^2\Big[I^-(w,0)-I^-(w,-2c) -I^+(w,0)+I^+(w,-2c)\cr
&-&I^-(w,2c)+I^-(w,0)+I^+(w,2c)-I^+(w,0)\Big]
\end{eqnarray}
so that, taking into account the symmetry properties of the
integrals,  $E_M^{\downarrow \uparrow}$ is found to vanish. In
conclusion, the magnetostatic energy difference $\Delta E_M$ turns
out to be just the opposite of the magnetostatic self-energy of
the single-domain prism. Its explicit expression is given in the
main text, Eq.~(\ref{explicit}), in terms of the in-plane aspect
ratio, $p=a/c$, and of the reduced thickness, $t=b/c$, of the
rectangular prism.

\end{document}